# Cu metal / Mn phthalocyanine organic spinterfaces atop Co with high spin polarization at room temperature

*E. Urbain, F. Ibrahim, M. Studniarek, F. Ngassam, L. Joly, J. Arabski, F. Scheurer, F. Bertran, P. Le Fèvre, G. Garreau, E. Denys, P. Wetzel, M. Alouani, E. Beaurepaire, S. Boukari, M. Bowen\*, W. Weber\**

E. Urbain, Dr. M. Studiarnek, F. Ngassam, Dr. L. Joly, J. Arabski, Dr. F. Scheurer, Prof. M. Alouani, Dr. E. Beaurepaire, Dr. S. Boukari, Dr. M. Bowen, Prof. W. Weber
Institut de Physique et Chimie des Matériaux de Strasbourg
Université de Strasbourg, CNRS UMR 7504
23 rue du Loess, BP 43, F-67034 Strasbourg Cedex 2 (France)
E-mail: martin.bowen@ipcms.unistra.fr, wolfgang.weber@ipcms.unistra.fr
Dr. F. Bertran, Dr. P. Le Fèvre
Synchrotron SOLEIL, L'Orme des Merisiers, Saint-Aubin, BP 48, 91192 Gif-sur-Yvette, France
Dr. G. Garreau, Dr. E. Denys, Prof. P. Wetzel
Institut de Science des Matériaux de Mulhouse,
CNRS-UMR 7361, Université de Haute-Alsace, 68057 Mulhouse, France

**Abstract**

The organic spinterface describes the spin-polarized properties that develop, due to charge transfer, at the interface between a ferromagnetic metal (FM) and the molecules of an organic semiconductor. Yet, if the latter is also magnetic (e.g. molecular spin chains), the interfacial magnetic coupling can generate complexity within magnetotransport experiments. Also, assembling this interface may degrade the properties of its constituents (e.g. spin crossover or non-sublimable molecules). To circumvent these issues, one can separate the molecular and FM films using a less reactive nonmagnetic metal (NM). Spin-resolved photoemission spectroscopy measurements on the prototypical system Co(001)//Cu/Mn-phthalocyanine (MnPc) reveal that the Cu/MnPc spinterface atop ferromagnetic Co is highly spin-polarized at room temperature, up to Cu spacer thicknesses of at least 10 monolayers. Ab-initio theory describes a spin polarization of the topmost Cu layer after molecular hybridization that can be accompanied by magnetic hardening effects. This spinterface's unexpected robustness paves the way for 1) integrating electronically fragile molecules within organic spinterfaces, and 2) manipulating molecular spin chains using the well-documented spin-transfer torque properties of the FM/NM bilayer.

The field of organic spintronics, which blends spin electronics with organic electronics[1] has recently received considerable attention[2-11] toward developing a new generation of spin devices for information technology applications. In contrast to the limited number of inorganic semiconductor candidates, the large number of available molecules can promote novel multifunctional devices that utilize exotic molecular properties such as spin crossover.[12] While molecular engineering conditions the fundamental electronic and optical properties of molecular films, deploying these properties within multifunctional (including spintronic) devices also requires appropriately engineering the interfaces between these molecules and metals.

A promising ingredient to achieve efficient organic spintronic devices is to utilize a ferromagnetic (FM) metal-organic interface. Indeed, first reports that spin-polarized charge transfer at this interface can magnetize the molecules' $3d$ sites[2,3] were later augmented by experimental evidence that the conduction states across these interfaces are highly spin-polarized[8] at room temperature (RT).[13-15] This notably accounts for very high spintronic performance (tunnel magnetoresistance (TMR)>3000 % at 2 K) upon elastic tunneling between these organic spinterfaces,[16] and constitutes a generic property proceeding from the adsorption of organic material onto a FM surface.[13-15] Note that this low density of highly spin-polarized interface states (IS) at the Fermi level $E_F$ can help resolve issues of resistivity mismatch[17] toward efficient spin injection into organic semiconductors.[18]

Unfortunately, the strong spin-polarized hybridization that accounts for the spinterface formation generates several challenges. 1) The congruent magnetic coupling at the FM/molecule organic spinterface can stabilize, at RT, the intrinsic correlations in the magnetic fluctuations within a molecular spin chain,[19] as demonstrated for the Co/MnPc bilayer[10]. This magnetic coupling generates complexity in understanding solid-state tunneling experiments across CoPc and $H_2$Pc/MnPc nanojunctions.[20,16] 2) Adsorbing molecules directly onto the FM surface severely limits the class of possible molecules and FM materials used. Indeed, wet deposition for instance would oxidize a transition metal FM electrode. Only a small subset of molecules can be sublimed under ultrahigh vacuum conditions required to maintain the FM surface properties. Finally, the large charge transfer upon molecular adsorption onto a transition metal surface can degrade a molecule's fragile electronic properties such as spin crossover.[21] At present, it does not seem feasible to assemble a FM metal/organic molecule spinterface that combines high spin polarization with the additional spin crossover functionality.

One solution to these challenges is to decrease molecular hybridization by adsorbing the molecules onto a less reactive nonmagnetic (NM) layer placed between the FM and molecular layers. This solution exhibits several conceptual advantages. 1) The charge transfer resulting from molecular adsorption onto a NM surface can still be sizeable enough to magnetize the NM layer[22,23] through an effect called magnetic hardening. 2) An electronically fragile molecular property such as spin crossover can be maintained upon molecular adsorption onto a suitable noble NM layer.[24] 3) The underlying Co layer sets the NM/molecule interfacial magnetization through interlayer exchange coupling (IEC). In turn, the FM/NM bilayer's well-documented[25] spintronic properties (e.g. spin transfer torque) can potentially be utilized to study magnetotransport across the molecular spin chain, partly thanks to the reduced coupling between the FM layer and molecular spin chain here. This requires, however, that the IEC be sufficiently robust to stabilize[10] the molecular spin chain's exchange interactions.

Overall, while numerous studies deal with organic semiconductors directly deposited onto a FM substrate (see references above), only very little work on indirect magnetic coupling of organic semiconductors exists.[26,11] In particular, recent x-ray magnetic circular dichroism (XMCD) measurements have shown that the magnetism of Mn within Mn phthalocyanine (MnPc) molecules is coupled to the buried FM Co(001) layer across a separation layer of Cu.[11] However, these measurements could not evidence an induced magnetization[13] and/or spin polarization at the Fermi level, as predicted by DFT calculations,[11] on the other, numerous C and N sites that compose the interface. Furthermore, the physical mechanism of IEC, which was proposed to explain the spinterface formation, potentially places severe constraints on an industrial implementation of the effect. Indeed, this mechanism requires an ultrathin Cu layer with thickness control on the monolayer (ML)-level. In addition, magnetic coupling at RT was only found for very low (up to 4 ML) Cu coverage.

As an echo to the crucial advances in knowledge, regarding the FM/molecule organic spinterface, that were 1) magnetization on the molecule's $3d$ site[2,3] followed by 2) high interfacial spin polarization[8,13-16], we expand knowledge of the NM/molecule spinterface atop FM beyond the magnetization of the single $3d$ site of the molecule[11] with a study of the spin polarization around the Fermi level resulting from all interface's atomic sites. We performed spin-polarized photoemission spectroscopy (SpinPES) measurements at room temperature of the Cu/MnPc interface atop Co. While the molecule's $3d$ site magnetization is stabilized by IEC to the FM and thus oscillates with NM thickness[11], we find that the sign of the spin polarization near $E_F$ at the NM/molecule spinterface atop FM is independent of NM thickness. Furthermore, this spin polarization is high at RT, and remains high even for NM thicknesses reaching 10 ML. *Ab-initio* calculations show that interfacial hybridization involves the molecule's C *2p* states with spin-polarized Cu $3d$ states at the Cu/MnPc spinterface atop Co, notably around $E_F$. This could be associated with magnetic hardening effects[22,23].

We present in Fig. 1 (a-c) the spin-resolved electron distribution curves as a function of the electron energy $E-E_F$ for the three consecutive steps of the Co(001)/Cu/MnPc stack fabrication. When studying only the Co(001) surface (Fig. 1(a)), we observe a strong imbalance of the spin-up and spin-down channels resulting from the FM Co layer, which reflects the strong negative spin polarization P of Co close to the Fermi level $E_F$ (Fig. 1 (d)). This is because Co is a strong ferromagnet, i.e. without spin-up *d*-electrons at $E_F$. When Co is covered by 2.7 ML of Cu (Fig. 1 (b)), the spin-polarized signal is attenuated in both spin channels due to Cu. Additional photoemission intensity coming from Cu results in an overall reduction of the spin polarization (Fig. 1 (d)). The weak bump at about -0.8 eV energy in the spin-down channel is due to a negatively spin-polarized Cu quantum-well state (QWS; see SI).

Figure 1 (c) shows the photoemission signal for Cu(001)//Co/Cu covered by 0.8 ML MnPc. We observe the appearance of two structures: at around -0.25 eV in the spin-up channel, and at around -0.8 eV in the spin-down channel (both indicated in Fig. 1 (c)). These two features also appear in the spin polarization spectrum (Fig. 1 (d)). Against the backdrop of the Co-induced negative polarization, which is attenuated due to overlayer coverage, there appears around -0.25 eV energy a relative decrease in this negative polarization. Since this molecule-induced upturn is much stronger than in the neighboring energy ranges, we infer that the presence of the molecules is not merely suppressing the negative polarization of Co, but is yielding an additional positive polarization. We note that a similar reasoning was applied in the case of molecules directly deposited onto a ferromagnet.[13-15] To strengthen this point further, the data shown in the inset of Fig. 1 (d) for 4 ML Cu evidence a clear sign change of the spin polarization at the -0.25 eV energy position of the IS. This is only possible if the IS is highly positively polarized, and can by no means be explained by the appearance of an unpolarized IS. Turning now to the IS at -0.8 eV, its negative spin polarization is undoubtedly revealed by the fact that the total spin polarization (Fig. 1(d)) changes sign from positive to negative around -0.8 eV energy.

To more directly evidence the highly polarized IS, we examine spin-polarized difference spectra so as to focus on the spin-resolved photoemission signal that arises from both the interface and the molecular layers atop the interface. To extract this signal, we use a subtraction procedure that takes into account the molecule-induced attenuation of the Co/Cu substrate photoemission signal. This procedure and its justification are detailed in Refs. [13-15]. Figures 2 (a-c) show the energy dependence of spin-polarized difference spectra at RT for Co//Cu/MnPc stacks with three different Cu thicknesses (1.3, 2.7, and 7 ML). The intensity close to the Fermi level is dominated by a spin-up structure whose maximum is, depending on the Cu thickness, between -0.2 and -0.4 eV energy. The spin-down channel, on the other hand, exhibits a small gap. Consequently, the spin polarization of the interfacial electronic band structure is strongly positive close to the Fermi level for all three Cu thicknesses, as expected from the energy-dependence of the spin polarization of the total photoemission signal discussed previously (Fig. 1(d)). We surmise that the IS at -0.8 eV is more difficult to resolve within the difference spectra due to a much larger line width compared to that of the IS close to $E_F$. This is compounded by an overall increase in background intensity in both spin channels due to molecular states.

For the measurement with 10 ML Cu thickness, we refrain - because of the reduced signal-to-noise ratio - from determining the spin-polarized difference spectra, and show instead in Fig. 2(d) the energy dependence of the spin polarization both for the Co/Cu reference and the molecule-covered stack. As seen for lower Cu thicknesses (see discussion of Fig. 1 (d)), we also witness for 10 ML Cu coverage a strong molecule-induced upturn of the polarization at about -0.2 eV. We thus infer that the spin polarization of the interfacial electronic band structure is strongly positive close to the Fermi level.

These RT experimental SpinPES measurements at the Cu/MnPc interface atop Co are surprising in two respects. 1) The spectral signature of the Cu/MnPc organic spinterface atop FM Co is very similar to that of the Co/MnPc spinterface. 2) It endures up to at least 10 ML Cu spacer thickness. In what follows, we address these points.

It is astonishing that the SpinPES results obtained here with a Cu interlayer between the MnPc molecules and the FM Co are similar to those obtained with MnPc directly deposited onto Co.[13] To exclude any possibility of a direct contact between the molecules and the Co film, we utilized the extreme surface sensitivity of ion-scattering spectroscopy (ISS) to determine Cu ML coverage completion. The inset to Fig. 1(b) shows the evolution of the normalized Co peak area in an ISS experiment versus Cu thickness. The Co concentration at the surface decreases rapidly with Cu coverage: it is as low as 10 % for 1 ML of Cu, and negligible at 1.3 ML Cu. We can thus exclude any direct contact between the molecules and the Co(001) film in our SpinPES experiments with at least 1.3 ML Cu.

When organic molecules directly adsorb onto a ferromagnet's surface, it is believed that the IS results from the hybridization of the FM $3d$ states with the $2p$ states of the molecule (e.g. Refs. [7,27]), with a comparatively small contribution from a $3d$ molecular site if present.[6] Figure 3(a) shows calculated spin-dependent $3d$ density-of-states (DOS) of the top Cu layer in the system Co(3ML)/Cu(3ML) (i.e. without molecule). Spectral features with relatively a strong spin polarization are found over the whole energy range studied. Importantly, although much weaker compared to the DOS for energies below -1.5 eV (not shown), a significant spin-polarized $d$ DOS is also evidenced at the bare Cu surface between -1 eV and $E_F$. Additional calculations show that this significantly spin-polarized Cu surface $d$ DOS is also present for larger Cu thicknesses (4 and 5 ML Cu on top of 3 ML Co; data not shown). This spin-polarized $d$ DOS exhibits only minor alterations when MnPc is adsorbed onto Co(3ML)/Cu(3ML) (see Figure 3(b), compare with panel (a)). Finally, Figure 3(c) shows the $p$ DOS of the MnPc layer on top of Co(3ML)/Cu(3ML). The data clearly reveals the existence of highly polarized $p$ states around the Fermi level and for lower energies. Vertical grey lines across these datasets clearly show how certain spin-polarized spectral features are common to the Cu $d$ states and MnPc $p$ states at the Cu/MnPc interface, notably around $E_F$, *i.e.* are involved in the formation of the Cu/MnPc spinterface atop Co.

The hybridization of the molecule's *p* states with spin-polarized Cu 3*d* states at the Cu surface, and their involvement within the spinterface formation, can thus explain the spin polarization of the IS at RT regardless of Cu spacer thickness up to at least 10 ML. As a note, QWS would contribute a negative spin polarization to that of the Cu/MnPc interface due to boundary conditions[28] in contrast to the positive spin polarization of the IS close to $E_F$. Moreover, the energy position of the QWS depends on the Cu spacer thickness, and is expected to cross the -0.2 to -0.4 eV energy range of the IS only for 4.5 and 9 ML Cu interlayer thickness,[29] yet we observe that the energy position of the spin-polarized IS is only weakly changing with Cu thickness. Nevertheless, experimental photoemission parameters such as the collection angle, photon energy and reciprocal space point can condition the strength of the QWS-induced modulation of spin polarization, including an eventual sign change.[30-32] We also emphasize that the oscillation in sign of magnetic coupling between the molecule's Mn 3*d* site and the Co layer[11] concerns the spinterface's entire DOS, while we are only considering here the spin-polarization of the DOS near $E_F$ within a limited *k* space around the Γ point. We thus conclude that 1) QWS cannot alone explain formation of the NM/molecule spinterface atop FM due to the thickness independent behavior of the IS; and 2) that this organic spinterface stems from features of the spin-polarized *d* DOS at the Cu surface of a Co/Cu bilayer whose alteration upon molecular adsorption could also involve magnetic hardening effects.[22,23] This origin of the spinterface formation in terms of a spin-polarized density of *d* states at a noble metal surface naturally extends to Au surfaces, which do not freeze the spin state of a spin crossover molecule upon adsorption.[24]

To summarize our work, by performing spin-resolved photoemission spectroscopy experiments on MnPc molecules separated from FM Co by a Cu interlayer, we have observed a highly spin-polarized organic spinterface at RT close to the Fermi level. The effect is only weakly changing with Cu thickness and persists at least up to 10 ML Cu spacer thickness. These experimental results are supported by our *ab-initio* calculations, which show that the interfacial Cu 3*d* DOS in the Cu/MnPc spinterface atop Co is significantly spin-polarized and hybridized with the MnPc's *p* states. Our work relaxes the industrially constraining requirements of ultrathin Cu with ML thickness control suggested by our previous attribution of IEC as a driver of the spinterface formation.[11] This reinforces the NM/molecule spinterface atop FM as a promising, industrially flexible path to assembling organic spinterfaces using a wider array of molecular and ferromagnetic metal candidates whose properties or deposition conditions otherwise appear orthogonal to achieving such interfaces. We emphasize in particular that electronically fragile molecular properties such as spin crossover can be maintained upon adsorption onto noble metal surfaces.[24] This will help affirm the full chemical engineering potential of organic spintronics.

Looking ahead, our work also opens exciting prospects into spin chain-driven device physics. Indeed, the demonstrated robustness of the NM/molecule organic spinterface atop FM suggests that it may also stabilize the correlations in the magnetic fluctuations[19] within a molecular spin chain atop this spinterface. By reducing the magnetic coupling between FM and molecule thanks to the NM spacer, electrical manipulation of the spinterface[33,34] and the spin chain may be more easily distinguished. Furthermore, one can utilize the well-documented[25] spin transfer torque properties of the FM/NM bilayer to more precisely understand the fundamental interaction between the spin-polarized current and the spin chain's excited states within a solid-state device (*e.g.* in a well-crafted operando experiment[35]), with important mid-term repercussions for our information and communications technology society.

**Methods**

The spin-resolved photoemission experiments were performed on the CASSIOPEE beamline at Synchrotron SOLEIL using horizontally polarized photons of 20 eV energy impinging upon the sample at 45º. Photoelectrons are acquired in normal emission geometry.

The energy resolution is 130 meV. The measurement of the spin polarization is done through a Mott detector, which exploits the left-right asymmetry A of electron scattering due to spin-orbit coupling.[36] The effective Sherman factor S of the Mott detector is 0.12. To eliminate any experimental asymmetry (e.g. due to a misalignment of the electron beam with respect to the Mott detector), photoemission spectra for opposite magnetization directions are measured. The spin polarization component P perpendicular to the scattering plane is given by P=A/S. The spin-up and spin-down intensity spectra are obtained, because of $P=(I_{up}-I_{down})/(I_{up}+I_{down})$, as follows: $I_{up}=(I/2)(1+P)$ and $I_{down}=(I/2)(1-P)$ with I the spin-integrated photoemission signal.

Prior to any deposition, the single-crystal Cu(001) substrate is cleaned by several cycles of Ar-ion sputtering and annealing at 800 K. Then, a FM film of Co is first deposited at room temperature from a Co rod heated by electron beam bombardment. We note that the film system Cu(001)//Co has been extensively investigated in the past (see Refs. [37-40]). As a second step, a Cu spacer layer is deposited onto the Co film at room temperature. We emphasize that this is a well-studied system exhibiting layer-by-layer growth.[41] The Cu thickness was calibrated by Reflection High Energy Electron Diffraction (RHEED). In the third step of the sample preparation, MnPc molecules are deposited by radiative heating onto the Cu film at room temperature. The evaporation rate is controlled by a quartz microbalance. The thicknesses of the molecular layers are determined by Auger electron spectroscopy. Scanning tunneling microscopy (STM) studies (see SI) show that the MnPc molecules remain intact when deposited onto a Cu(001) surface.

The density functional theory (DFT) calculations were carried out by means of the VASP package[42] and the generalized gradient approximation for exchange-correlation potential as parametrized by Perdew, Burke, and Ernzerhof.[43] We used the projector augmented wave (PAW) pseudopotentials as provided by VASP.[44] The van der Waals (vdW) weak interactions were computed within the so called GGA-D2 approach developed by Grimme[45] and later implemented in the VASP package.[46] A kinetic energy cutoff of 450 eV has been used for the plane-wave basis set.

**Acknowledgements**

We thank the SOLEIL staff for technical assistance and insightful discussions. We gratefully acknowledge support from the CNRS, the Institut Carnot MICA's 'Spinterface' grant, from ANR grant ANR-11-LABX-0058 NIE, from the Alexander-von-Humboldt Foundation and the Baden-Württemberg Stiftung in the framework of the Kompetenznetz für Funktionale Nanostrukturen (KFN), and from the International Centre for Frontier Research in Chemistry. Computations were performed using HPC resources from the meso-centre of Strasbourg and from GENCI-CINES Grant 2014-gem1100.

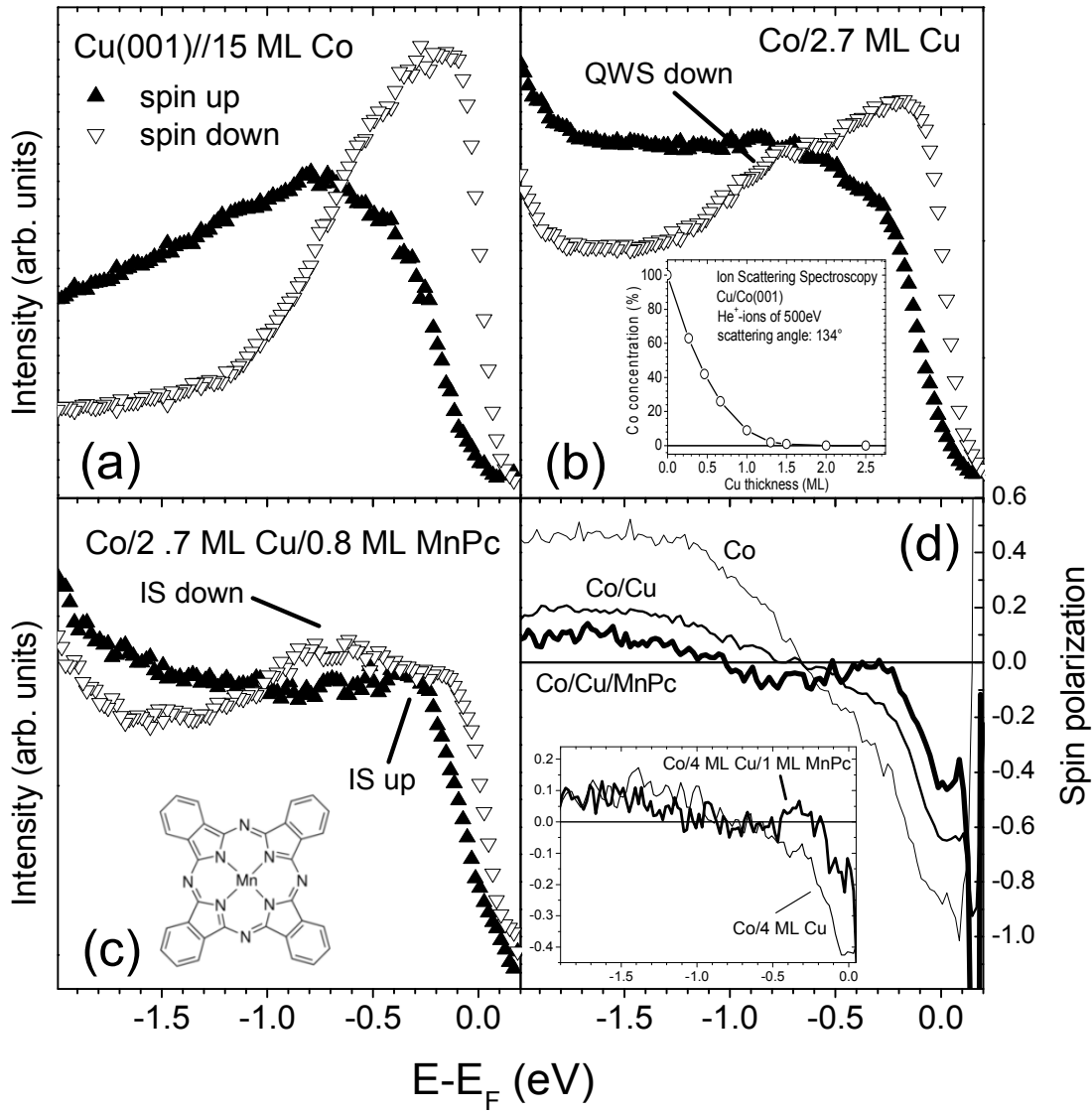

**Figure 1**: **Formation of the Cu/MnPc spinterface atop Co.** (a)-(c) Spin-resolved electron distribution curves as a function of the electron energy for consecutive steps of the Co/Cu/MnPc stack fabrication: (a) Cu(001)//15 ML Co, (b) Co/2.7 ML Cu, and (c) Co/2.7 ML Cu/0.8 ML MnPc. In (b) the presence of a Cu QWS in the spin-down channel is indicated. The appearance of IS due to MnPc coverage are indicated in (c). The inset to panel (b) shows the evolution of the normalized Co peak area in an ion scattering spectroscopy experiment versus Cu thickness. For Cu thicknesses above 1.3 ML the Co(001) film is completely covered by Cu. The inset in (c) shows a MnPc molecule. (d) Spin polarization at room temperature as a function of energy of the Co film, the Co/2.7 ML Cu film and of the same stack covered by 0.8 ML MnPc. The inset shows the spin polarization vs. energy of a Co/4 ML Cu film and of the same stack covered by 1 ML MnPc. Molecular adsorption onto Cu generates a positive net spin polarization against a negative, attenuated background of Co spin polarization.

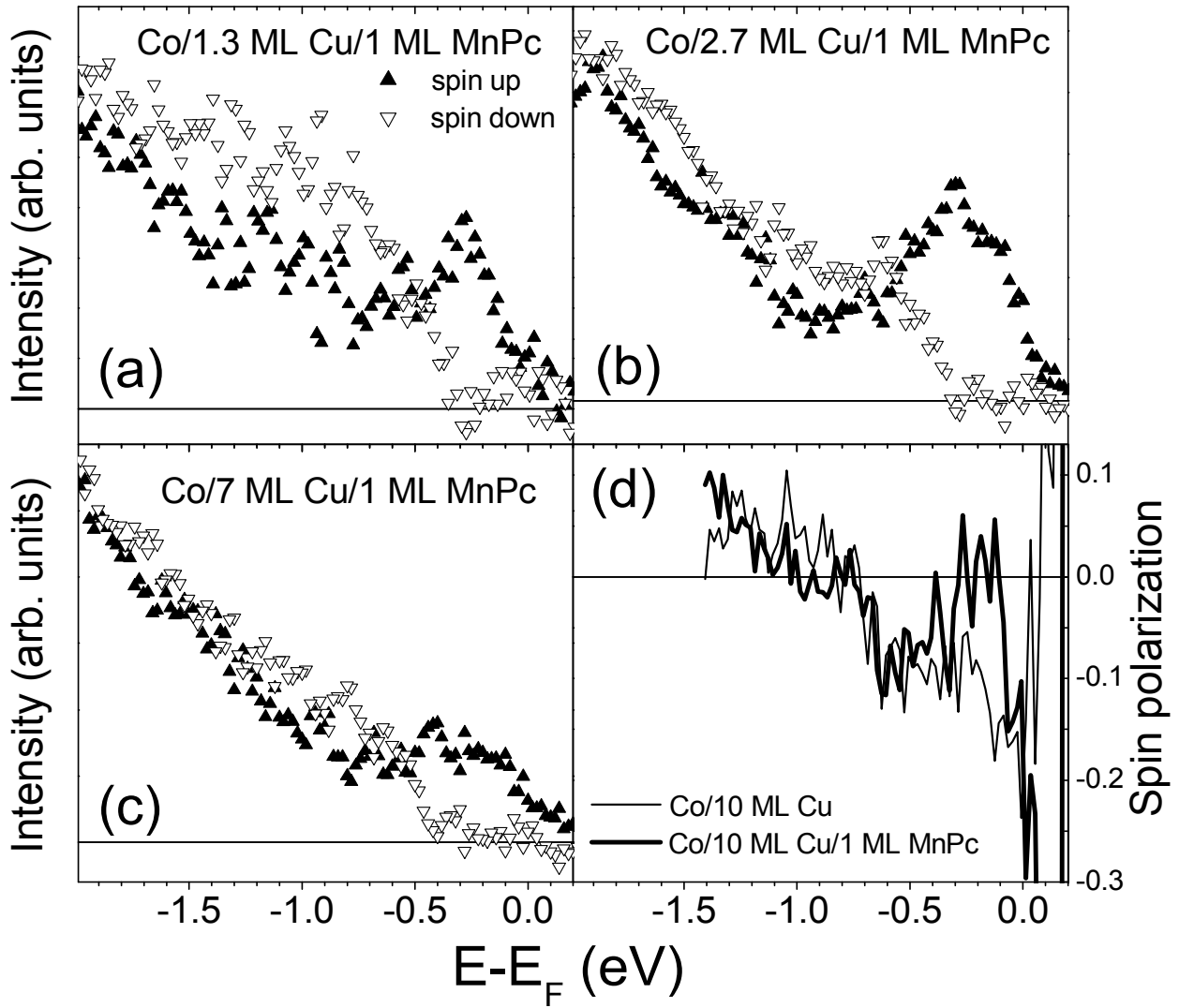

**Figure 2**: **High spin polarization of the Cu/MnPc spinterface atop Co.** (a)-(c) Spin-resolved electron distribution curves as a function of the electron energy arising from the Cu/MnPc spinterface, upon suitably subtracting spectra taken before molecular deposition from those acquired after (see Refs. [13-15] for details of the data analysis). (d) Spin polarization at room temperature as a function of the energy of the Co/10 ML Cu stack and of the same stack covered by 1 ML MnPc. The Cu/MnPc interface remains spin-polarized up to 10 ML Cu with only weak changes in spectral features upon changing Cu thickness.

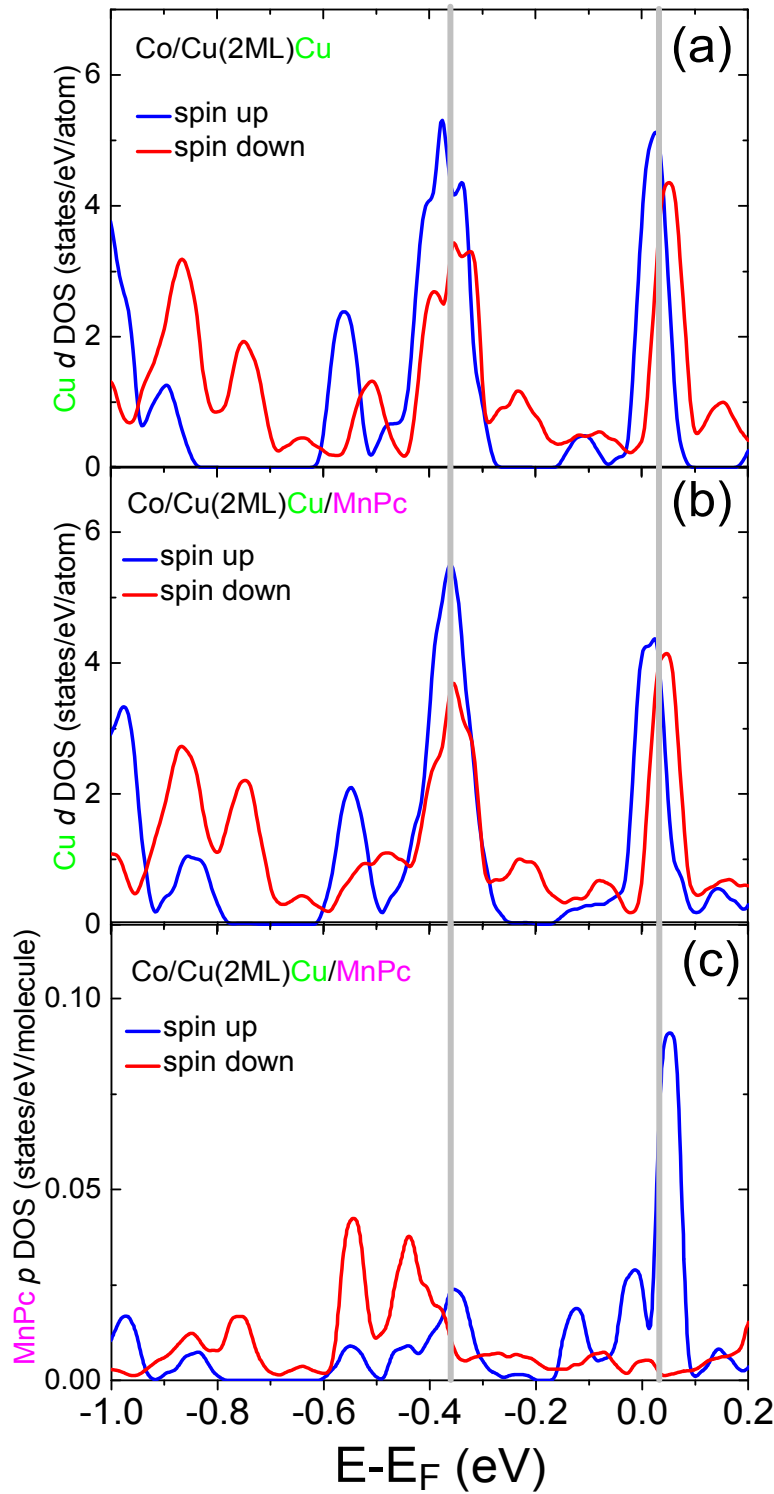

**Figure 3**: **Origin of the spinterface formation.** Spin-polarized density of the topmost Cu monolayer's *d* states for (a) Co(3ML)/Cu(3ML) and (b) Co(3ML)/Cu(3ML)/MnPc. (c) Spin-polarized DOS of the MnPc *p* DOS for Co(3ML)/Cu(3ML)/MnPc. All DOS data have been averaged over an energy interval of 50 meV. The adsorption of MnPc onto Co(3ML)/Cu(3ML) induces minor changes to the Cu surface's spin-polarized *d* DOS. Vertical gray lines indicate which spectral features are common to the topmost Cu monolayer's *d* states and to the MnPc's *p* states, i.e. contribute to the formation of the Cu/MnPc spinterface atop Co, notably around $E_F$.